# Magnetic phases evolution in the LaMn$_{1-x}$Fe$_x$O$_{3+y}$ system

O. F. de Lima

*Instituto de Física Gleb Wataghin, UNICAMP, 13083-970 Campinas, SP, Brazil*

J. A. H. Coaquira

*Núcleo de Física Aplicada, Instituto de Física, UnB, 70904-970 Brasília, DF, Brazil*

R. L. de Almeida, L. B. de Carvalho, S. K. Malik

*Centro Internacional de Física da Matéria Condensada, UnB, 70904-970 Brasília, DF, Brazil*

ABSTRACT

We have investigated the crystal structure and magnetic properties for polycrystalline samples of LaMn$_{1-x}$Fe$_x$O$_{3+y}$, in the whole range $0.0 \leq x \leq 1.0$, prepared by solid state reaction in air. All samples show the ORT-2 orthorhombic structure that suppresses the Jahn-Teller distortion, thus favoring a ferromagnetic (FM) superexchange (SE) interaction between Mn$^{3+}$-O-Mn$^{3+}$. For x = 0.0 the oxygen excess (y ≈ 0.09) produces vacancies in the La and Mn sites and generates a fraction around 18% of Mn$^{4+}$ ions ($t_{2g}^3$) and 82% of the usual Mn$^{3+}$ ions ($t_{2g}^3, e_g^1$), with possible double exchange interaction between them. The Fe doping in this system is known to produce only stable Fe$^{3+}$ ions ($t_{2g}^3, e_g^2$). We find an evolution from a fairly strong FM phase with a Curie temperature $T_C \sim 160$ K, for x = 0.0, to an antiferromagnetic (AFM) phase with $T_N =$ 790 K, for x = 1.0, accompanied by clear signatures of a cluster-glass behavior. For intermediate Fe contents a mixed-phase state occurs, with a gradual decrease (increase) of the FM (AFM) phase, accompanied by a systematic transition broadening for 0.2 < x < 0.7. A model based on the expected exchange interaction among the various magnetic-ion types, accounts very well for the $M_S$ dependence on Fe doping.

## I.  INTRODUCTION

The study of manganites started almost sixty years ago[1], with discussions on the occurrence of mixed-valence states and strong ferromagnetic (FM) interaction between the $Mn^{3+}$ and $Mn^{4+}$ ions, in the $(La_{1-x}Ca_x)MnO_3$ and $(La_{1-x}Sr_x)MnO_3$ systems. After the discovery of large magnetoresistance[2] and colossal magnetoresistance (CMR)[3] effects in manganite thin films, about fifteen years ago, interest on the study of these materials has been revived.

The stoichiometric parent manganite $LaMnO_3$ is an A-type antiferromagnetic insulator, with a Néel temperature ($T_N$) of 140 K. In this compound all Mn ions are trivalent and coupled through a superexchange (SE) interaction[4,5] that produces a FM coupling within the a-b planes in its orthorhombic structure, and an antiferromagnetic (AFM) coupling between these planes, along the c direction. When a fraction of trivalent La ions is substituted by divalent alkaline earth ions (e.g., Sr and Ca) the same fraction of $Mn^{3+}$ is transformed to $Mn^{4+}$, and an electron will be allowed to hop between these ions producing a FM double-exchange (DE) interaction[6], which also promotes a metallic electric conduction.

Much of the work done so far to explore manganite properties[7,8], has been mainly focused on the effects produced by doping the La sites, such as the lattice distortions that ultimately influence the DE and SE interactions. Also play an important role the combined effects derived from spin, charge, and orbital degrees of freedom, typical in strongly-correlated electron systems. The occurrence of a magnetic mixed-phase state in manganites, consisting of short-range-ordered regions that coexist with long-range-ordered regions, is by now strongly supported by many experimental[9,10,11,12] and theoretical[13,14] results. In that context the various possibilities of charge-, spin-, and orbital-ordered configurations play a central role.

Substitutions on the Mn site by various $3d$ cations (Cr, Co, Ni and Fe) have been explored[15,16,17,18,19,20,21,22] in the last years, mainly motivated by the fact that Mn is the main actor in the DE interaction, and also to test interesting ideas related to other possible coupling mechanisms between Mn ion and the doping cation. For instance, a ferromagnetic SE interaction between $Mn^{3+}$ - O - $Mn^{3+}$, has been proposed and actually verified in the system[23] $LaMn_{1-x}B_xO_{3+y}$ (B = Ga, Co, Ni). There is also the idea of a FM coupling caused by a DE interaction between $Fe^{3+}$ - O - $Mn^{3+}$ in the $LaMn_{1-x}Fe_xO_3$ system[19]. In this paper we will be presenting strong evidences for the FM coupling via SE interaction between $Mn^{3+}$ ions, for the case of $LaMn_{1-x}Fe_xO_{3+y}$. The substitution of $Fe^{3+}$ (S = 5/2) for $Mn^{3+}$ (S = 2) is special, in the sense that both these ions have the same ionic radius[7,24] and different magnetic moments in their high spin states.

This means that in this case lattice distortions can be avoided and any existing magnetic change has a good chance to be appropriately detected.

The first study on manganites with Mn substituted by Fe was done more than fifty years ago[25], in the $La_{0.85}Ba_{0.15}Mn_{1-x}Fe_xO_3$ system. A good explanation of the obtained resistivity data was based on the available ionic interactions, where the idea of a mixed-phase state was essential. Until recently all studies of Fe doping Mn sites were done in the usual CMR compounds La(Sr, Ca)$Mn_{1-x}Fe_xO_3$. Typically they found[17,18] that electric conduction and ferromagnetism were consistently suppressed by Fe doping, and CMR was shifted to lower temperatures being enhanced by Fe doping in some cases. However, doping the simpler parent compound $LaMnO_3$, could help to identify the real influence of Fe in the structure. This would avoid more complex effects related to structural and electronic modulations in the La-O layers. Therefore, recently some few works have been describing and discussing about the structural, transport and magnetic properties of the $LaMn_{1-x}Fe_xO_{3+y}$ system[19,20,21,22]. In general these studies agree with the suppression of ferromagnetism by increasing Fe doping, in reason of the gradual substitution of $Mn^{3+}$ by $Fe^{3+}$, that reduces the amount of DE-coupled $Mn^{4+}$ - O - $Mn^{3+}$ bonds. Also they agree on the occurrence of a cluster-glass-like behavior, with complicated features, mainly for higher Fe doping. However a study that covers the whole range of Fe doping, which could provide a comprehensive description of the phases evolution, is still lacking.

In this paper we present a thorough study of the crystal structure and magnetic properties for a relatively large set of $LaMn_{1-x}Fe_xO_{3+y}$ samples, with $0.0 \leq x \leq 1.0$. We found a suppression of ferromagnetism while $x$ increases, and a complex evolution of a mixed-phase state, with clear signatures for the occurrence of a cluster-glass (CG) in the whole doping range. A model that describes accurately the saturation magnetization of all samples, based on the evolution of possible magnetic phases, is proposed and discussed.

## II.  EXPERIMENTAL DETAILS

A set of nine samples of composition $LaMn_{1-x}Fe_xO_{3+y}$ , for $0.0 \leq x \leq 1.0$, were reacted using the solid diffusion method in open air. For each sample we started with the stoichiometric mixture of $La(OH)_3$, $Mn_2O_3$ and $Fe_2O_3$ powders, which was fired in a tubular furnace at the temperature of $1100^{o}C$, for 24 hours. Following, the pre-reacted sample was crushed and thoroughly milled with a mortar and pestle, then pressed in a cylindrical die and returned to the furnace for a reaction and sintering treatment, again at $1100^{o}C$ for 24 hours. After the samples were prepared, small pieces with masses in the range of 20 - 80 mg were cut from them for the magnetic measurements. Also small pieces were pulverized for the X-ray experiments, which were done in a Phillips diffractometer using Cu-K$_\alpha$ radiation.

It is well established that a perovskite structure cannot accept excess $O^{2-}$ ions in an interstitial site[7]. Actually it has been shown that any oxygen excess y, in the non-stoichiometric compound $LaMnO_{3+y}$, is accommodated as cation vacancies[26,27]. Therefore, from a crystallographic point of view a more appropriate formula would be $La_{1-z}Mn_{1-z}O_3$, where z = y/(3 + y), by assuming an equal amount of La and Mn vacancies[27]. In this paper however, for simplicity, we adopt the formula $LaMn_{1-x}Fe_xO_{3+y}$, considering that vacancies are equally distributed in the La and Mn sites.

Figure 1 shows the X-ray spectra for seven samples, covering the whole range of iron doping. All of them, as well as the samples with x = 0.05 and 0.50 (not shown), produced similar spectra, where the corresponding peaks fall essentially at the same angular positions. However for x > 0.7 a slight decrease in the angular distance between peaks is observed, revealing a small increase of the lattice spacing. These results are consistent with the well established fact that iron occupies the manganese sites always in the valence state[19,28] $Fe^{3+}$ that has essentially the same ionic radii of $Mn^{3+}$, whose value is[24] 0.645 Å. On the other hand, the $Mn^{4+}$ ionic radii is smaller ($\sim$ 0.530 Å)[24], thus causing some lattice expansion when a $Mn^{4+}$ site is occupied by a $Fe^{3+}$ ion.

Analysis of all diffraction patterns by the Rietveld method indicated that all samples contain at least 90% of pure manganite phase, with an orthorhombic structure (space group *Pnma*). The main impurities detected for x = 0.7 and x = 0.9 have their strongest peaks marked in Fig. 1 by a solid square ($La_2O_3$), an asterisk ($La(OH)_3$) and a down arrow ($\alpha$-$Fe_2O_3$). The calculated lattice parameters for the low end composition (x = 0.0) are a = 5.535 Å, b = 7.786 Å and c = 5.500 Å, within the expected values that characterize the so-called O-type orthorhombic structure (ORT-2)[29,30] where c < b/2$^{1/2}$ < a. This structure suppresses the Jahn-Teller distortion, which involves a cooperative rotation of the $MnO_6$ octahedras, and favors an isotropic FM SE-interaction[4] between $Mn^{3+}$ - O - $Mn^{3+}$. In contrast, the more usual O'-orthorhombic structure (ORT-1)[4,29,30,31], where b/2$^{1/2}$ < c < a, stabilizes the Jahn-Teller distortion and favors a canted AFM superexchange-interaction between $Mn^{3+}$ - O - $Mn^{3+}$. The ORT-1 structure is typically observed for $LaMnO_{3+y}$ with y < 0.05, when reacted in a reducing or oxygen-depleted atmosphere, while samples reacted in air have normally the ORT-2 structure[7,29,30], with y $\approx$ 0.09. Since in our work all reactions were processed in air, we then assume an oxygen excess of y = 0.09 in all samples, following the typical values reported by different authors.

Table I lists all samples and several of their magnetic properties, extracted from a large number of magnetization curves that were taken with a Quantum Design SQUID magnetometer.

Zero field cooled (ZFC) and field cooled (FCC) measurements were done in almost all cases. In order to warrant the same initial conditions for the magnet residual field and for the samples magnetic state, before each ZFC curve a field of 5 kOe was applied and lowered to zero through oscillations, at T = 300K. Then the temperature was lowered to 2K, where the measuring field was applied and the measurements were taken during a slow warming ramp, up to 320 K. Following, the FCC curve was taken in a cooling ramp, using the same rate, down to 2K. Ac susceptibility measurements were taken with a Physical Property Measurement System (PPMS from Quantum Design).

## III. RESULTS AND DISCUSSIONS

Typical magnetization curves for all samples, under a reasonably small applied field of 100 Oe, are shown in Fig. 2(a). Clearly there is a ferromagnetic-like (FM-like) transition that looks sharp for small iron contents (x ≤ 0.1). However this transition becomes broader and is shifted to lower temperatures, showing smaller intensity, for higher x values. Indeed, one can see in the inset of Fig. 2(a) that even for x = 1.0, the high end composition also exhibits a clear, although weak, FM-like signature. Since LaFeO$_3$ is known to present an AFM coupling with spin canting[32], we believe this is possibly the origin for the observed weak ferromagnetism. Another relevant feature present in the M×T curves is the strong irreversibility between the ZFC and FCC curves, typical of a superparamagnetic (SPM) relaxation phenomena of a cluster-spin-glass (CG) system[33,34,35,36]. In fact several features, which will be highlighted along this paper, support the likely occurrence of a CG in our LaMn$_{1-x}$Fe$_x$O$_{3+y}$ samples, similar to results for some compositions of this system reported by other authors[21,37]. For instance, all of our ZFC curves show a maximum at a temperature T$_p$ (around 150 K for x = 0.0) below the FM-like transition, and a shoulder at a lower temperature T$_f$ (around 100 K for x = 0.0) that fades out for higher x values. Fig. 2(b) shows some ac susceptibility measurements for sample x = 0.0 confirming the occurrence of these especial temperatures, by showing a frequency dependent (independent) maximum at T$_f$ (T$_p$) in the imaginary components, accompanied by a corresponding inflexion points in the real components. Several characteristics of T$_p$ and T$_f$, whose details will be published elsewhere, typify T$_p$ as an average blocking temperature where the clusters moments begin to freeze in a field H, and T$_f$ as the temperature where this thermally activated freezing process reaches a maximum. Consistent with that picture there is a strong field dependence, revealed by a complete suppression of the magnetic irreversibility down to T = 2 K, for relatively small values of H, as can be seen in Fig. 4 for H = 3 kOe. This might be due to a full alignment of the clusters moments in the magnetic field direction, as already observed in LaMnO$_{3+y}$ samples[38] with y ≤ 0.15. Similar observations were also reported for nanoparticles of γ-Fe$_2$O$_3$ incorporated in a resin matrix[39], as well as for amorphous Pd-Ni-Fe-P alloys[40].

Figure 3 shows plots of the virgin magnetization curves as a function of $H/T$, for the sample with x = 0.3. These curves are spaced from each other by 3 K, in the interval from 26 K to 68 K. One sees that five curves for T ≥ 56 K superimpose almost perfectly, thus following what is predicted to happen for SPM spin-clusters[41], while for T < 56 K the curves gradually break away. A complementary test is shown in the inset of Fig. 3 where an Arrott plot[42], $M^2 \times H/M$, is performed for the same set of isothermal curves. In this standard experimental method the occurrence of FM order is predicted to occur when straight lines $M^2 \propto H/M$ are obtained in the plots. Further, it defines the Curie temperature (T$_C$) of the isotherm whose linear extrapolation intercepts the vertical axis at the value zero. In the present case we find T$_C$ ≈ 56 K (black dots in Fig. 3) for the sample with x = 0.3, in excellent agreement with the temperature value that limits the SPM behavior, according to the first test of Fig. 3. We conclude, then, that a SPM regime occurs for T > T$_C$, while a FM-like order is established for T ≤ T$_C$. It is worth noticing that the down curvatures in the Arrott's plots, at low fields and especially for the lower temperature isotherms, are usually observed in granular and amorphous ferromagnets[40,43]. The T$_C$ values listed in Table I were defined at the inflexion point where the derivative dM/dT has a maximum value (see Fig. 2(a)). This is a simpler and commonly used criterion, although it furnishes T$_C$ values which are slightly above those obtained through Arrott plots. In this work the T$_C$ values coincide with the point where the ZFC and FCC curves bifurcate, for samples having x ≤ 0.2. For samples with higher iron contents T$_C$ is located below the bifurcation point, thus indicating a more complex dynamics of the SPM clusters.

We assume that a short-range order FM-like state is induced by the applied magnetic field, in the region between the superparamagnetic and cluster-glass states ($T_f < T < T_C$), following the same interpretation applied to similar data obtained with amorphous alloys[40,44]. In our work three facts give support to this hypothesis: i) the relatively high and positive values of the calculated Weiss constant ($\theta_W$) (Table I); ii) the Arrott plots typical of FM materials (inset of Fig. 3); and iii) the clear separation between the SPM and FM-like regimes shown in the scaling plot of Fig. 3. However it is not clear at the moment what would be the microscopic mechanism that could provide such FM-like order among the SPM clusters.

Figure 4 shows a plot of the inverse susceptibility $H/M$ as a function of temperature for seven samples, measured under $H$ = 3 kOe in ZFC and FCC modes. For $0.0 \leq x \leq 0.2$, a large part of the higher-temperature region follows a linear behavior as predicted for a PM phase according to the Curie-Weiss law, $M/H = C/(T - \theta_W)$, where $C$ and $\theta_W$ are the Curie and Weiss constants, respectively. Therefore, the slope of a straight line fitted in the linear region (solid lines in Fig. 4) gives $1/C$ and its intercept with the temperature axis gives $\theta_W$. Since the effective magnetic moment per formula unit (f. u.) can be expressed, in Bohr magnetons ($\mu_B$), by[45]

$\mu_{eff} = (3k_B CA/N_A)^{1/2} \approx 2.829(AC)^{1/2}$, where $k_B$ is the Boltzmann constant, $A$ is the molecular weight and $N_A$ is the Avogadro number, one can see from Fig. 4 that $\mu_{eff}$ in general decreases when $x$ increases. But the experimentally determined $\mu_{eff} \approx 5.9$ $\mu_B$/(f. u.) for sample x = 0 (Table I) exceeds the spin-only value given by $g[S(S+1)]^{1/2}$, with $g = 2$, even if we assume the over estimated situation of having all manganese ions $Mn^{3+}$ in their high-spin state (S = 2), which would give 4.9 $\mu_B$/(f. u.) for sample x = 0.0. The other extreme with x = 1.0 is more complicated, since $LaFeO_{3+y}$ is expected to be in a canted-AFM state up to $T_N$ with no room, in principle, for a PM state when T < 790 K. Unless, of course, if an existing cluster system, as already assumed, could explain the transition from a FM-like state to a PM-like state, by increasing the temperature. However, we are not aware of any model that could provide a calculation of $\mu_{eff}$ in such a cluster system, for the case of x > 0.0 in our samples.

The well behaved trend, just described above, changes for iron contents $x \geq 0.3$, where two regions become gradually evident in Fig. 4 as x increases. In the first region, right above the curved section between the FM-like to SPM states, is observed a linear behavior (solid lines) that maintains more or less the same trend as observed in samples with x ≤ 0.2, where $\mu_{eff}$ and $\theta_W$ decrease gradually while x increases. Concomitantly, the second region bends to the right and departs progressively from the initial linear behavior (dashed lines). Other studies[28,46], on the single composition $LaMn_{0.5}Fe_{0.5}O_3$, also revealed this intriguing curvature in their H/M data, whose origin was attributed to the occurrence of magnetic clusters. We agree with that interpretation and, indeed, our data support a broader and more complete description of this phenomenon, based on the global magnetic evolution manifested by the whole set of samples (0.0 ≤ x ≤ 1.0).

The temperature $T_d$, indicated in the curve for x = 0.7 (Fig. 4), has been identified[40] as a transition point that separates a linear PM region, corresponding to the usual single magnetic moment per formula unit, from a curved SPM region, where the number of aligned moments gradually increases inside the clusters down to $T_C$. We partially agree with that interpretation, the only difference being that in our case a situation with a single moment per formula unit was not observed, although a PM-like Curie-Weiss description seems to work well.

In an attempt to shed more light on the problem, we now discuss the M×T curves of samples with x = 0.5, 0.7 and 1.0 (Fig. 5), taken in the range of temperatures between 300 K – 800 K and under a magnetic field of 5 kOe. A clear AFM transition occurs for the end compound LaFeO$_{3+y}$ at $T_N \approx$ 790 K, which is larger than the value of 740 K found in Mössbauer studies[32]. The strong irreversibility between the ZFC and FCC curves, again, indicates the occurrence of a CG system, plausibly formed by weak-FM domains or clusters of the canted AFM phase[32]. For the samples x = 0.5 and 0.7 the ordering transition around 790 K is not convincingly resolved, possibly due to a combination of two factors, the smaller relative content of the LaFeO$_{3+y}$ compound and a limited sensitivity of the measuring technique. However it is very interesting to observe the gradual increase of irreversibility, consistent with the progressive formation of LaFeO$_{3+y}$ as x increases. Notice also that these M×T curves are shifted vertically by 0.2 emu/g and 0.1 emu/g, as indicated by the vertical arrows in Fig. 5, aimed at solely to improve the clarity of presentation.

The inset in the upper part of Fig. 5 displays an enlarged view of the ZFC curve for LaFeO$_{3+y}$. In fact it matches very nicely with the end of the ZFC magnetization curve measured between 2 K – 320 K (not shown). However the monotonical decrease of M(T) that starts above $T_C$ = 65 K (see inset of Fig. 2(a)) is smoothly changed to a monotonical increase around 430 K, speeding up very quickly above 750 K. Without discussing the details of the underlying CG dynamics, we can conclude that this result must certainly affect the Curie-Weiss behavior observed in the first region of Fig. 4 (solid lines). Therefore, we believe this could be the origin of the unusual bending to the right, observed in the second region of Fig. 4 (dashed lines). This would account naturally for the gradually larger curvatures occurring for higher x values, which could be ascribed simply to the stronger presence of LaFeO$_{3+y}$. In other words, one could think that right above the first region starts a convoluted regime that combines the Curie-Weiss behavior, connected with the SPM region at low temperature, with some other functional dependence that accounts for the high temperature behavior of LaFeO$_{3+y}$.

Magnetization curves of complete M×H loops were measured for all samples, starting at H = 0 and then cycling between 50 kOe and -50 kOe, at the temperatures of 2 K (Fig. 6(a)) and 300 K (Fig. 6(b)). The curves for samples with x ≤ 0.1 are almost reversible presenting very small coercivities, $H_{C,300K}$ < 10 Oe at 300 K and $H_{C,2K}$ < 100 Oe at 2 K (see Table I and Fig. 7), common in systems of almost non-interacting SPM clusters or particles. However for x ≥ 0.2 a visible hysteresis appears at low fields and coercivity gradually increases. Curiously for x ≥ 0.7 coercivity values seems to stay around 2 kOe when measured at 2 K, in contrast with a steep increase, that reaches 17.5 kOe for sample x =1.0, when measured at 300 K.

The saturation magnetization at 2 K ($\mu_{H, 2K}$) was determined at the ordinate point intercepted by the linear extrapolation from the straight high-field region of M×H. This criterion eliminates the linear non-saturating behavior which is increasingly visible for samples with x ≥ 0.2, due probably to the growing presence of canted-AFM clusters of $LaFeO_{3+y}$. Consistent with that hypothesis, one sees in Fig. 6(a) that $\mu_{H, 2K}$ can definitely attain its saturated values for magnetic fields around 30 kOe, when x ≤ 0.1.

The canting angle for AFM ordered $LaFeO_3$ was determined through Mössbauer studies[32] to be $\alpha = \mu_{FM} / 2\mu_{Fe} \approx 0.009$ rad, where $\mu_{FM}$ is the measured canted FM component and $\mu_{Fe}$ is the $Fe^{3+}$ sublattice moment, assumed to be 5 $\mu_B$ (S = 5/2). This $\alpha$ value was also verified to be practically constant in a broad temperature interval, from $T_N$ = 740 K down to the lowest measured T ≈ 70 K. The measured value[32] $\mu_{FM} \approx 0.09$ $\mu_B$ is in remarkable agreement with $\mu_{H, 2K}$ = 0.07 $\mu_B$ that was measured in our x = 1.0 ($LaFeO_{3+y}$) sample, as indicated in Table I.

We also calculated the saturation magnetization at 300 K ($\mu_{H,300K}$), using the same procedure employed at 2 K, and their values (Table I) are about two orders of magnitude smaller than the $\mu_{H, 2K}$ values. The inset of Fig. 6(b) is a magnified view near H = 0 that shows an increasing hysteretic behavior for x ≥ 0.2. Since at 300 K only the $LaFeO_{3+y}$ clusters must present magnetic order in the sample (neglecting any small amounts of magnetic impurities), we indeed expect that $\mu_{H,300K}$ should increase with x as found. Interestingly its maximum value of 0.014 $\mu_B$, for x = 1.0, is only about 20% of the fully aligned canted-spins moments of $LaFeO_{3+y}$ measured at 2K. This is probably due to the temperature induced disorder of the canted-AFM clusters, since the canting angle itself is known to be temperature independent[32] in $LaFeO_3$.

The high coercivity values found in our x = 1.0 sample is very common in orthoferrites[45], where typical crystal anisotropies ($K_a$) of the order of $10^4$ erg/cm³ are combined with very small saturation magnetization values, usually smaller than 8 emu/cm³. High coercivity values in M×H curves similar to ours were found[47] in bulk polycrystalline (8.4 kOe) and amorphous (11.9 kOe) $Mn_3O_4$, which is a canted ferrimagnet below $T_C \approx 42$ K. It is worthwhile to notice that a coercivity value of 1170 Oe has also been reported[28] for a sample of $LaMn_{0.5}Fe_{0.5}O_3$ at 12 K, close to the value of 1160(10) Oe measured in our sample with x = 0.5, at 2 K (see Table I). Attributing the origin of coercivity only to crystal anisotropy, we can estimate from our data $K_a$ = $\mu_{H, 2K}$ $H_{C,2K}$ /2 ≈ 7×$10^2$ erg/cm³, at 2 K, and $K_a$ = $\mu_{H, 300K}$ $H_{C,300K}$ /2 ≈ 1.5×$10^3$ erg/cm³, at 300 K. The smaller value found at 2 K suggests another origin of coercivity. For instance, it could be that inter-clusters rotation dominates at low temperatures, while intra-cluster spin inversion dominates at high temperatures.

The inset of Fig. 7 shows a plot of the transition width, $\Delta T_{SPM} = T_d - T_C$, corresponding to the temperature-width of the main SPM region (see Fig. 4) as a function of x, evaluated from measurements taken under H = 100 Oe. It is reasonable to expect that a distribution of cluster sizes in this region, which evolves from smaller to larger sizes[11,40], might affect the magnetic transition width from a PM-like region (T > Td) to a FM-like region (T < $T_C$). Very interestingly the end compositions, x ≤ 0.1 and x ≥ 0.9, show relatively narrower transitions when compared to the central region, 0.2 ≤ x ≤ 0.7. This is a reliable result, even taking in consideration the large error bars involved in these calculations. A natural explanation for this behavior seems to be the occurrence of a magnetic mixed phase region in the intermediate compositions, discussed in more detail in the next section, while near the end compositions a more uniform cluster size distribution is established.

## IV.   THE MAGNETIC MIXED-PHASE MODEL

Figure 8 shows graphically most of the parameters listed in Table I which characterize the samples magnetic behavior. The effective magnetic moment ($\mu_{eff}$) and saturation moments ($\mu_{H, 2K}$, $\mu_{H, 300K}$) refer to the right ordinate axis in units of $\mu_B$/(f.u.), while the Curie ($T_C$) and Weiss ($\theta_W$) temperatures refer to the left axis. Notice that the evaluated error bars more or less coincide or are smaller than the symbols size in all cases. $\mu_{eff}$ starts with a value close to 5.9 $\mu_B$ at $x = 0.0$, which is higher than the overestimated theoretical value of 4.9 $\mu_B$ expected for a spin-only contribution of $Mn^{3+}$ ions. As $x$ increases, $\mu_{eff}$ goes down following an oscillating path and reaching a value of 3.1 $\mu_B$/(f.u.) at $x =1.0$, too high for $LaFeO_{3+y}$ in its canted-AFM state[32]. All these features possibly happen due to the presence of SPM clusters, as already discussed. $T_C$ and $\theta_W$ also goes down while $x$ increases, with $\theta_W > T_C$ up to $x =0.7$, characteristic of a FM coupling. For $x \geq 0.9$ there is an inversion, $T_C > \theta_W$, that can be attributed[4] to the dominance of AFM interactions.

The saturation magnetization measured at 2K ($\mu_{H, 2K}$) displays a linear decrease, varying from 3.75 (± 0.01) $\mu_B$ to 1.02 (± 0.02) $\mu_B$, when $x$ goes from 0.0 to 0.30. Then, it suddenly changes the decreasing rate for x > 0.30, still showing roughly a linear behavior up to x = 1.0. Following, we present a simple model, based on the magnetic ions evolution expected for the whole range of Fe doping, which can provide an excellent description for the $\mu_{H, 2K}$ behavior. A basic general assumption of this model is that iron atoms are incorporated into the $LaMn_{1-}$ $_x$Fe$_x$O$_{3+y}$ samples in the form of Fe$^{3+}$ ions[17,28], whose amount grows linearly up to 100%, by occupying the Mn$^{3+}$ and Mn$^{4+}$ ion sites.

The solid lines in Fig. 9 shows in more detail that the initial concentration of 18% for Mn$^{4+}$ is assumed to be invariant up to x = 0.82, where the total amount (82%) of Mn$^{3+}$ sites becomes fully occupied. Above that concentration the Mn$^{4+}$ sites are then steadily occupied by the Fe$^{3+}$ ions up to x = 1.0. This assumption is motivated by the fact that Mn$^{3+}$ ions have the same size[7] (0.645 Å) of Fe$^{3+}$ ions, while the Mn$^{4+}$ ions are smaller (0.530 Å), thus requiring a higher activation energy to be replaced by the Fe$^{3+}$ ions. The saturation magnetization ($M_S$), represented by the solid lines running very near to the $\mu_{H, 2K}$ points in Fig. 8, was calculated by the following equations, in units of $\mu_B$:

$$M_S = 4(0.82 - x) + 3(0.18) - 5x \quad ; \quad \text{for} \quad 0 \le x \le 1/3 \tag{1}$$

$$M_S = 4(0.82 - x) + (1.28 - 2.22x) + (4.79x - 3.26) \quad ; \quad \text{for} \quad 1/3 \le x \le 0.82 \tag{2}$$

$$M_S = 3(x - 1) + 3.7(1 - x) \quad ; \quad \text{for} \quad 0.82 \le x \le 1.0 \tag{3}$$

These equations consider the saturation moment aligned with the magnetic field H to be 4 $\mu_B$ for Mn$^{3+}$ (S = 2), 3 $\mu_B$ for Mn$^{4+}$ (S = 3/2) and 5 $\mu_B$ for Fe$^{3+}$ (S = 5/2), with all these ions in their high spin states. The first region is dominated by FM couplings and goes from x = 0.0 up to the special concentration x = 1/3. This value is identified as the limit for the higher rate of FM coupling suppression of both types, Mn$^{3+}$ - O - Mn$^{4+}$ or Mn$^{3+}$ - O - Mn$^{3+}$. Up to this concentration, each Fe$^{3+}$ makes an AFM coupling with a neighbor Mn$^{3+}$ or Mn$^{4+}$ and contributes with a magnetic moment of 5 $\mu_B$ opposite to the H direction, as expressed in Eq. (1). It is interesting to notice that all parameters plotted in Fig. 8 suffer an abrupt change at x ≈ 1/3, presenting slower variations above that point, consistent with the increasing dominance of the AFM order. The last region (0.82 ≤ x ≤ 1.0) is simple to analyze, since it admits only AFM couplings of two types, Fe$^{3+}$ - O – Fe$^{3+}$ or Fe$^{3+}$ - O – Mn$^{4+}$. This latter type most possibly allows the spins to flip over, in order to save magnetic energy, by aligning the larger moment of the Fe$^{3+}$ ions parallel to H. Therefore, the term 3(x − 1) in Eq. (3) describes a linear suppression of the antiparallel moment for the full content (18%) of Mn$^{4+}$, going from  - 0.54 $\mu_B$ to zero. Concomitantly the net moment for a corresponding fraction of 18% Fe$^{3+}$ varies from 3.7 $\mu_B$ to zero, as described by the second term of Eq. (3). Here we must notice that this effective moment, smaller than the full value of 5 $\mu_B$ for Fe$^{3+}$, was obtained from a fit to the $\mu_{H, 2K}$ data. Finally, the region 1/3 ≤ x ≤ 0.82 is perhaps the more complex, showing at the same time a decrease of FM and increase of AFM couplings, with the occurrence of spin flips in the latter. By increasing $x$ the average rate of FM suppression falls with respect to the first region, because part of the $Fe^{3+}$ ions will be occupying $Mn^{3+}$ sites that already belonged to some pre-existing AFM coupling of the type $Mn^{3+}$ - O - $Fe^{3+}$, thus becoming $Fe^{3+}$ - O - $Fe^{3+}$, which is still AFM but with a null contribution to the total magnetic moment. The last term in parenthesis of Eq. (2) describes the evolution for the $Fe^{3+}$ contribution to the total magnetic moment, by assuming that its effective moment is 4.79 $\mu_B$ in the whole region. This value was found by requiring that the total $Fe^{3+}$ moment must be equal to the border values, -5/3 $\mu_B$ at $x = 1/3$ and 2/3 $\mu_B$ at $x = 0.82$. This simple phenomenological approach considers that, by connecting linearly the intermediate region with the two well established contiguous regions, a proper account for the combined FM suppression and spin flips in the newly formed AFM bonds becomes guaranteed. Still in the same approach, the middle term in parenthesis of Eq. (2) accounts solely for the gradual spin flip of the invariant amount of $Mn^{4+}$ ions, whose total magnetic moment goes from 0.54 $\mu_B$ (at x = 1/3) to - 0.54 $\mu_B$ (at x = 0.82). The first terms in both Eqs. (1) and (2), simply describe the linear decrease of the $Mn^{3+}$ contribution to the FM couplings in the first region, and to the FM as well as AFM couplings in the second region.

Figure 9 also sketches another possible model, where the $Mn^{3+}$ and $Mn^{4+}$ ions evolution are represented by dash-dotted lines. Following a similar approach employed in the previous more accurate model the saturation magnetization was calculated and the result is represented by the dashed lines in Fig. 8. Clearly the agreement with the $\mu_{H, 2K}$ data is not good for $x \geq 1/3$, although the general trend is maintained. In particular, the small slope change of the data around x = 0.82 could not be captured, since this simplistic model assumes a linear decrease of the manganese ions ($Mn^{3+}$ and $Mn^{4+}$) content in the whole range $0.0 \leq x \leq 1.0$.

Finally it is worth mentioning that both models predict $M_S = 0$ for x = 1.0, while the experimental result is $\mu_{H, 2K} = 0.07(2)$ $\mu_B$. This happens because the models do not take into account the very small intrinsic contribution coming from the canted-AFM component of $LaFeO_{3+y}$, as well as eventual contributions arising from possible traces of magnetic impurities.

## V. CONCLUSION

We prepared a set of nine samples of polycrystalline $LaMn_{1-x}Fe_xO_{3+y}$, with iron doping uniformly distributed in the whole range $0.0 \leq x \leq 1.0$, using a solid diffusion reaction method. X-ray diffraction data and Rietveld analysis indicated good quality of samples containing at least 90% of the manganite phase, all them having orthorhombic structure (space group *Pnma*). Since the samples were reacted in air we estimated and oxygen excess of y = 0.09 in all samples, and the occurrence of ORT-2 type of orthorhombic structure, where c < b/2$^{1/2}$ < a, with lattice parameters a = 5.535 Å, b = 7.786 Å and c = 5.500 Å, for the sample with $x$ = 0.0.

Magnetization and ac susceptibility measurements, allowed a thorough characterization of the magnetic properties at low (2 K – 320 K) and high temperatures (300 K – 800 K). In general, the Curie and Weiss temperatures ($T_C$ and $\theta_W$), as well as the effective moment ($\mu_{eff}$) and the saturation moments at 2 K and 300 K ($\mu_{H, 2K}$ and $\mu_{H, 300K}$) decrease, while iron doping ($x$) increases. This was interpreted as an evolution of the magnetic phases that starts in x = 0.0 with a FM behavior, due to $Mn^{3+}$ - O - $Mn^{4+}$ double-exchange and $Mn^{3+}$ - O - $Mn^{3+}$ superexchange couplings, and evolves to a FM-like behavior, mainly caused by a canted-antiferromagnetism originated from a gradual increase of $Fe^{3+}$ - O - $Fe^{3+}$ bonds. These bonds belong to the orthoferrite LaFeO$_{3+y}$, the end composition at x = 1.0, that shows a Néel temperature ($T_N$) of 790 K. It is worth noticing that very high coercivities ($H_C$ ~ 18 kOe), typical of orthoferrites, were indeed obtained for x = 1.0.

All samples showed a clear cluster-spin-glass behavior, revealed by several features like, e.g., the occurrence of irreversible M×T curves consistent with the frequency-dependent peak positions in ac susceptibility curves, and a strong field-dependence of these irreversibilities, possibly due to the alignment of the cluster's magnetic moments. Several results also suggest the occurrence of a SPM behavior in all samples, such as the excellent overlapping of virgin magnetization curves for temperatures above $T_C$, when plotted against the scaling variable $H/T$, and $\mu_{eff}$ values much larger than the theoretical spin-only predictions. This is especially evident in the high iron doping region (x ≥ 0.7) where only a small magnetic moment would be expected from a canted-antiferromagnetic LaFeO$_{3+y}$. Following previous works, we hypothesize that a short-range order (FM-like state) could eventually be induced by the applied magnetic field, in the region between the superparamagnetic and cluster-spin-glass states.

A magnetic mixed-phase model was proposed, by starting with a basic general assumption that iron atoms are incorporated into the LaMn$_{1-x}$Fe$_x$O$_{3+y}$ structure in the form of $Fe^{3+}$ ions, whose amount grows linearly up to 100%, by occupying the $Mn^{3+}$ and $Mn^{4+}$ ion sites. The estimated initial concentration of 18% for $Mn^{4+}$ ($x$ = 0.0) was assumed to be invariant up to $x$ = 0.82, when the total amount (82%) of $Mn^{3+}$ sites becomes fully occupied. Above that concentration the $Mn^{4+}$ sites are then steadily occupied by the $Fe^{3+}$ ions up to x = 1.0. By considering that all ions are in their high-spin states, $Mn^{3+}$ (S = 2), $Mn^{4+}$ (S = 3/2) and $Fe^{3+}$ (S = 5/2), the model allowed an accurate description of the $\mu_{H, 2K}$ data in the whole range of iron doping.

## ACKNOWLEDGMENTS

We acknowledge the financial support from the Brazilian science agencies FAPESP (Fundação de Amparo à Pesquisa do Estado de São Paulo ), CNPq (Conselho Nacional de Desenvolvimento Científico e Tecnológico) and CAPES (Coordenação de Aperfeiçoamento de Pessoal de Nível Superior). One of us (JAHC) wants to acknowledge also FINATEC (Fundação de Empreendimentos Científicos e Tecnológicos). We also thank Prof. A. Ferraz and Prof. S. Quezado for the interest in this work.

CAPTIONS        ( O. F. de Lima et al.  )

Table I – Magnetic properties of the $LaMn_{1-x}Fe_xO_{3+y}$ samples, calculated from magnetization measurements taken with a SQUID magnetometer. $T_C$ and $\theta_W$ are the Curie and Weiss temperatures, respectively; $\mu_{eff}$ is the effective moment in the PM state; $\mu_{H, 2K}$ and $\mu_{H,300K}$ are the saturation moments at 2 K and 300 K; $H_{C,2K}$ and $H_{C,300K}$ are the coercivities at 2 K and 300 K. The magnetic moments are given in units of Bohr magnetons per formula unit.

FIG. 1. X-ray diffraction patterns of all $LaMn_{1-x}Fe_xO_{3+y}$ samples, measured with Cu-K$_\alpha$ radiation. Numbers in parenthesis near the peaks of sample with x = 0.0 are the Miller indices corresponding to the main crystallographic planes. The strongest peaks of impurities are marked with a solid square ($La_2O_3$), an asterisk ($La(OH)_3$) and a down arrow ($\alpha$-$Fe_2O_3$), being detected mainly in samples with x $\geq$ 0.5.

FIG. 2. (a) ZFC and FCC magnetization curves of all $LaMn_{1-x}Fe_xO_{3+y}$ samples, measured with H = 100 Oe. The inset is a magnified view that helps to observe the FM-like transitions for samples with x $\geq$ 0.7; (b) real (left axis) and imaginary (right axis) components of magnetic ac susceptibility, measured in the sample with x = 0.0, at different frequencies.

FIG. 3. The main frame shows a scaling plot of the virgin magnetization as a function of H/T, for sample with x = 0.3, showing the collapse of five isothermal curves from 56 K to 68 K, suggesting a SPM behavior. The inset is an Arrott's plot for the same set of isotherms that varies from 26 K to 68 K, in steps of 3 K, revealing a Curie temperature of 56 K.

FIG. 4. Plot of the inverse susceptibility $H/M$ as a function of temperature, measured under $H$ = 3 kOe in ZFC and FCC modes. A simple Curie-Weiss behavior (solid lines) is seen for x $\leq$ 0.2, while for x $\geq$ 0.3 the curves bend to the right after a limited interval of straight-line behavior.

FIG. 5. High temperature magnetization curves of samples with $x$ = 0.5, 0.7and 1.0, measured in the range of temperatures between 300 K - 800 K and under H = 5 kOe. A clear AFM transition is seen for $x$ = 1.0 ($LaFeO_{3+y}$) at $T_N \approx$ 790 K. The inset is an enlarged view of the ZFC curve for $LaFeO_{3+y}$, showing a monotonical decrease that changes to a monotonical increase around 430 K, speeding up very quickly above 750 K.

FIG. 6. Magnetization as a function of field for all samples, measured at 2 K (a) and 300 K (b). Magnetization is expressed in Bohr magnetons per formula unit in (a), and S values indicate the high-spin states. For samples with $x \geq 0.2$, hysteresis gradually increases with $x$ in both temperatures. The inset in (b) magnifies a region near the origin, making clear the manifestation of high coercitivies for $x \geq 0.9$.

FIG. 7. Coercivity as a function of iron doping $x$, calculated from M×H curves taken at 2 K and 300 K. The inset shows a plot of $\Delta T_{SPM} = T_d - T_C$, which is the temperature width of the main SPM region (see Fig. 4) as a function of $x$, evaluated from measurements taken with H = 100 Oe.

FIG. 8. Most of the magnetic properties listed in Table I is plotted as a function of iron doping $x$. The effective magnetic moment ($\mu_{eff}$) and saturation moments ($\mu_{H, 2K}$, $\mu_{H, 300K}$) refer to the right ordinates in units of Bohr magnetons per formula unit, while the Curie ($T_C$) and Weiss ($\theta_W$) temperatures refer to the left ordinates. The vertical dashed lines mark the special concentrations x = 1/3 and x = 0.82 (see text). The horizontal dash-dotted lines indicate special spin-only values of $\mu_{eff}$ given by $2[S(S+1)]^{1/2}$.

FIG. 9. Mixed-phase model of the ion concentration evolution in the system $LaMn_{1-x}Fe_xO_{3+y}$, as a function of iron doping $x$. $Fe^{3+}$, $Mn^{3+}$ and $Mn^{4+}$ are assumed to be the only magnetic ions available in the system. The solid lines refer to a more accurate model; while the dashed-dotted lines refer to a simpler and less accurate model (see text).

| x | $T_C$ (K) | $\theta_W$ (K) | $\mu_{eff}$ ($\mu_B$/f.u.) | $\mu_{H, 2K}$ ($\mu_B$/f.u.) | $\mu_{H, 300K}$ ($10^{-2}$ $\mu_B$/f.u.) | $H_{C,2K}$ (Oe) | $H_{C,300K}$ (Oe) |
|---|---|---|---|---|---|---|---|
| 0.00 | 160(1) | 189(2) | 5.88(2) | 3.75(1) | 0.030(3) | 35(1) | 7(1) |
| 0.05 | 153(1) | 175(2) | 5.91(1) | 3.30(1) | 0.199(3) | 26(1) | 12(1) |
| 0.10 | 133(1) | 150(2) | 6.16(1) | 2.96(2) | 0.303(3) | 87(1) | 8(1) |
| 0.20 | 110(2) | 130(2) | 5.74(1) | 1.94(2) | 0.395(4) | 425(4) | 18(1) |
| 0.30 | 62(3) | 113(2) | 4.41(1) | 1.02(2) | - | 1900(20) | 80(2) |
| 0.50 | 65(3) | 111(2) | 4.53(3) | 0.61(3) | 0.520(4) | 1160(20) | 170(5) |
| 0.70 | 50(3) | 85(3) | 3.77(3) | 0.25(3) | 0.557(4) | 1790(20) | 470(10) |
| 0.90 | 43(4) | 22(4) | 4.02(5) | 0.08(3) | 0.648(5) | 2900(20) | 8000(100) |
| 1.00 | 65(3) | 40(2) | 3.10(5) | 0.07(2) | 1.448(5) | 1800(20) | 17500(200) |

Table I  -  ( O. F. de Lima et al. )

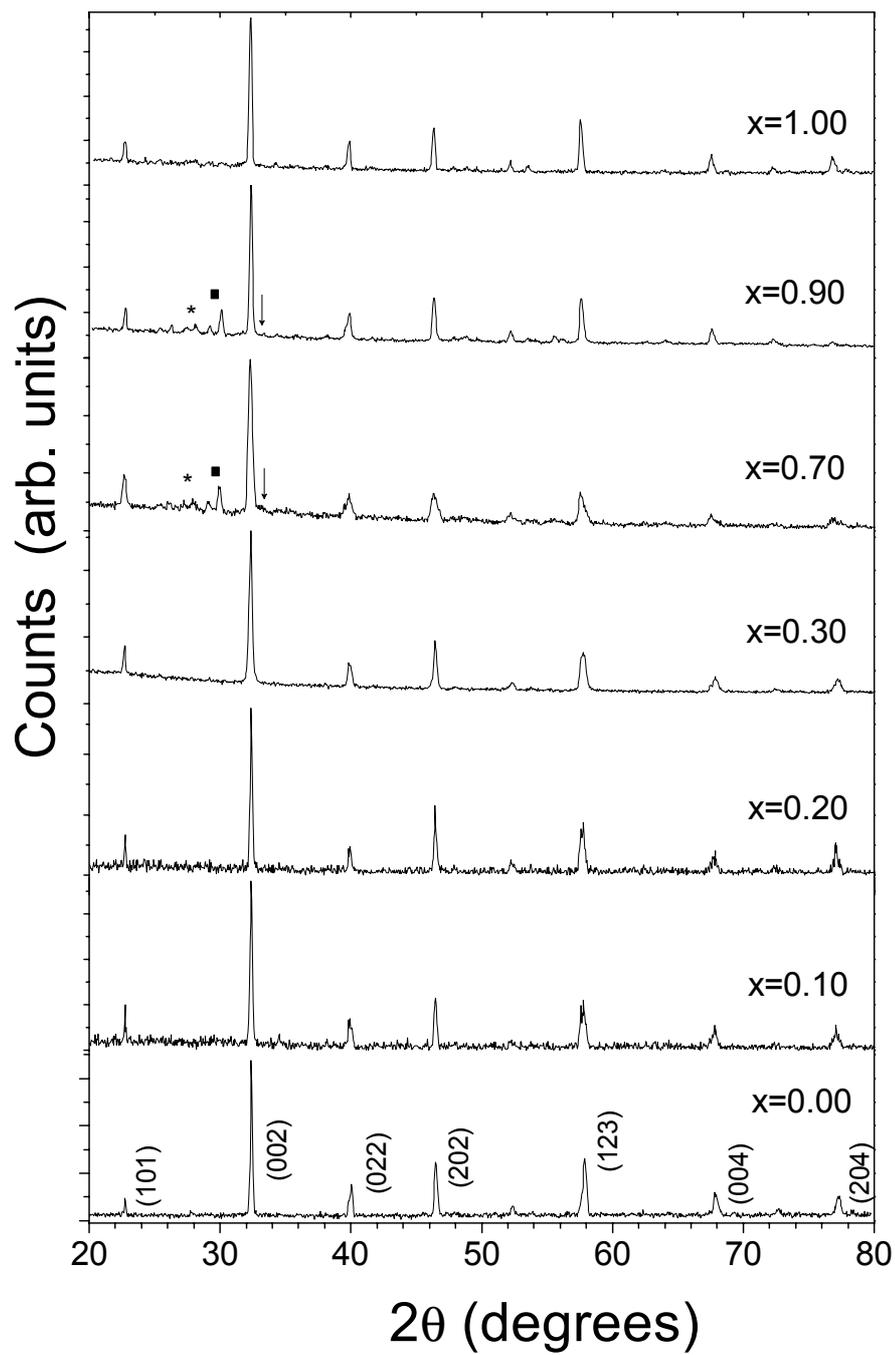

Counts (arb. units)

2θ (degrees)

x=1.00
x=0.90
x=0.70
x=0.30
x=0.20
x=0.10
x=0.00

(101)  (002)  (022)  (202)  (123)  (004)  (204)

Fig.1 - O. F. de Lima et al.

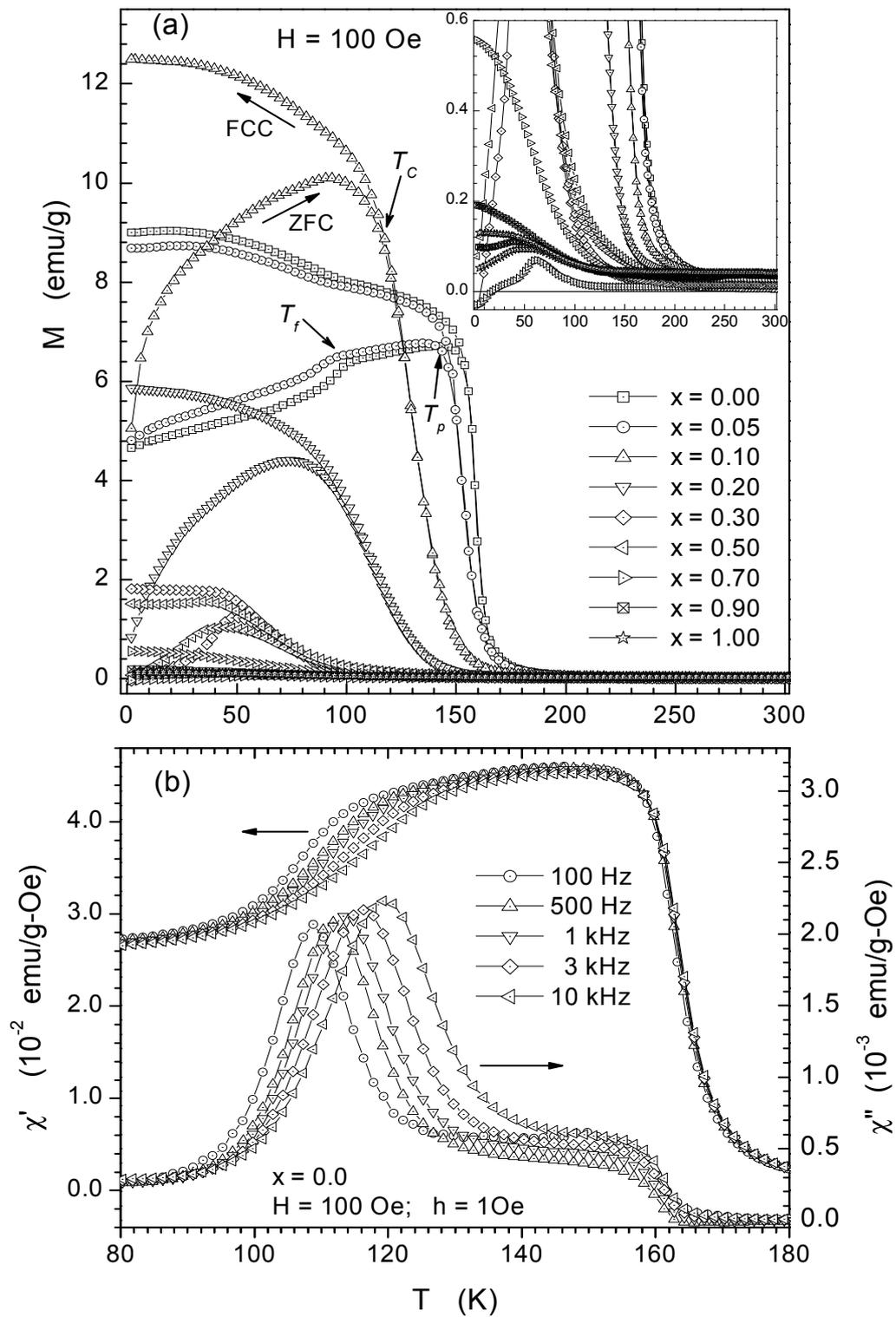

Fig.2 - O. F. de Lima et al.

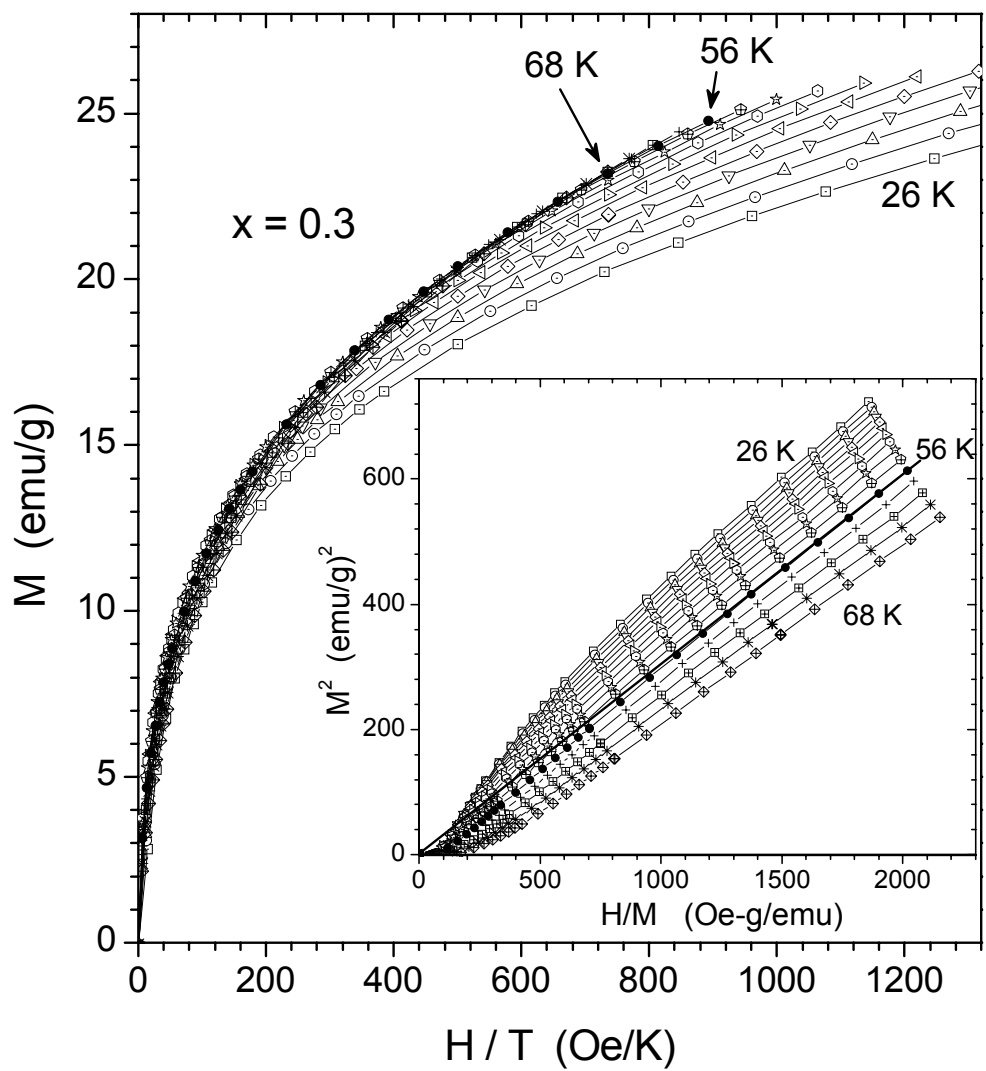

Fig.3 - O. F. de Lima et al.

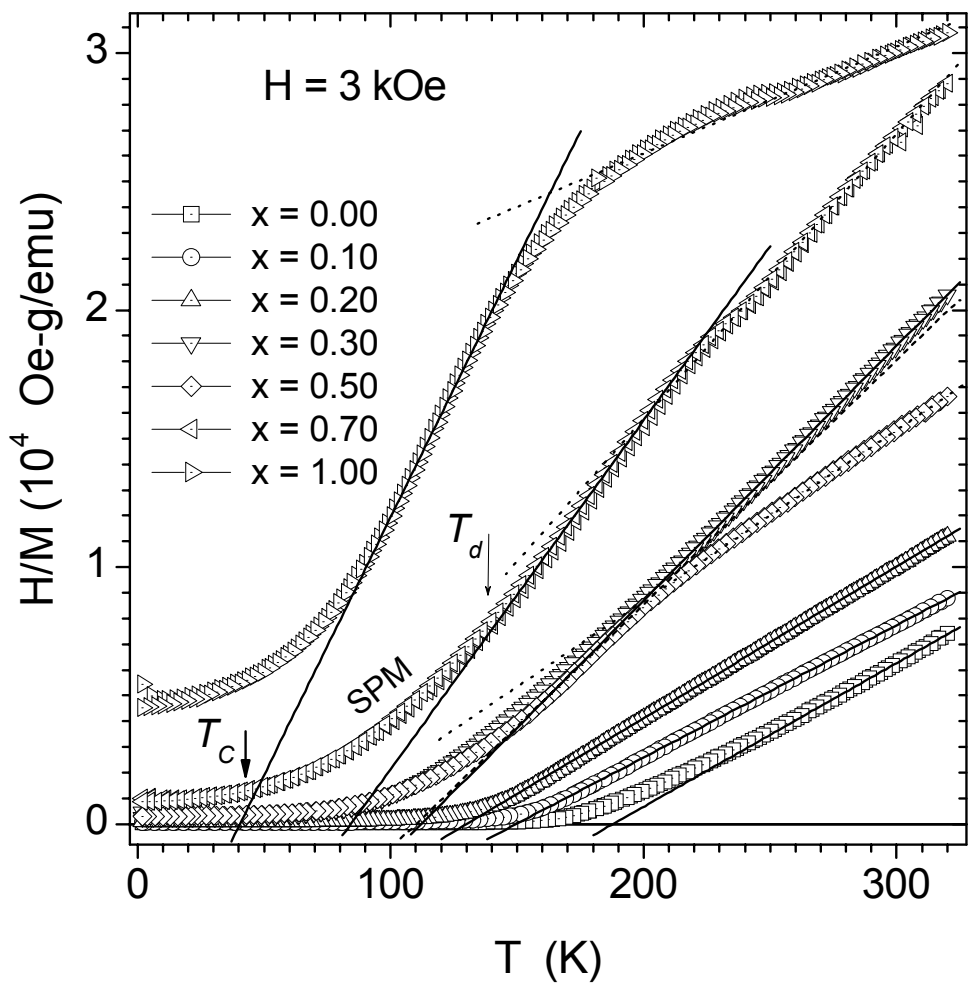

Fig.4 - O. F. de Lima et al.

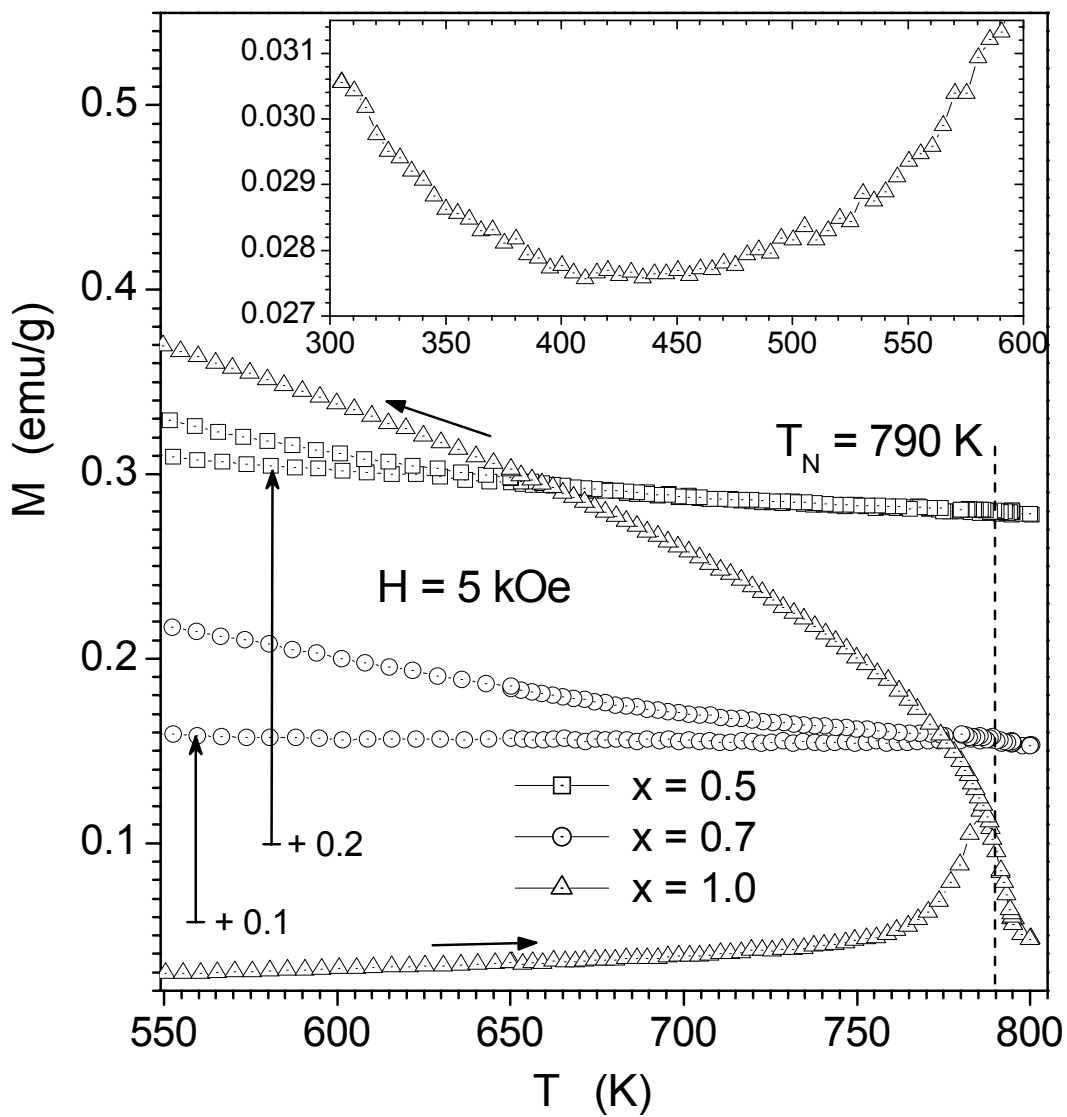

Fig.5 - O. F. de Lima et al.

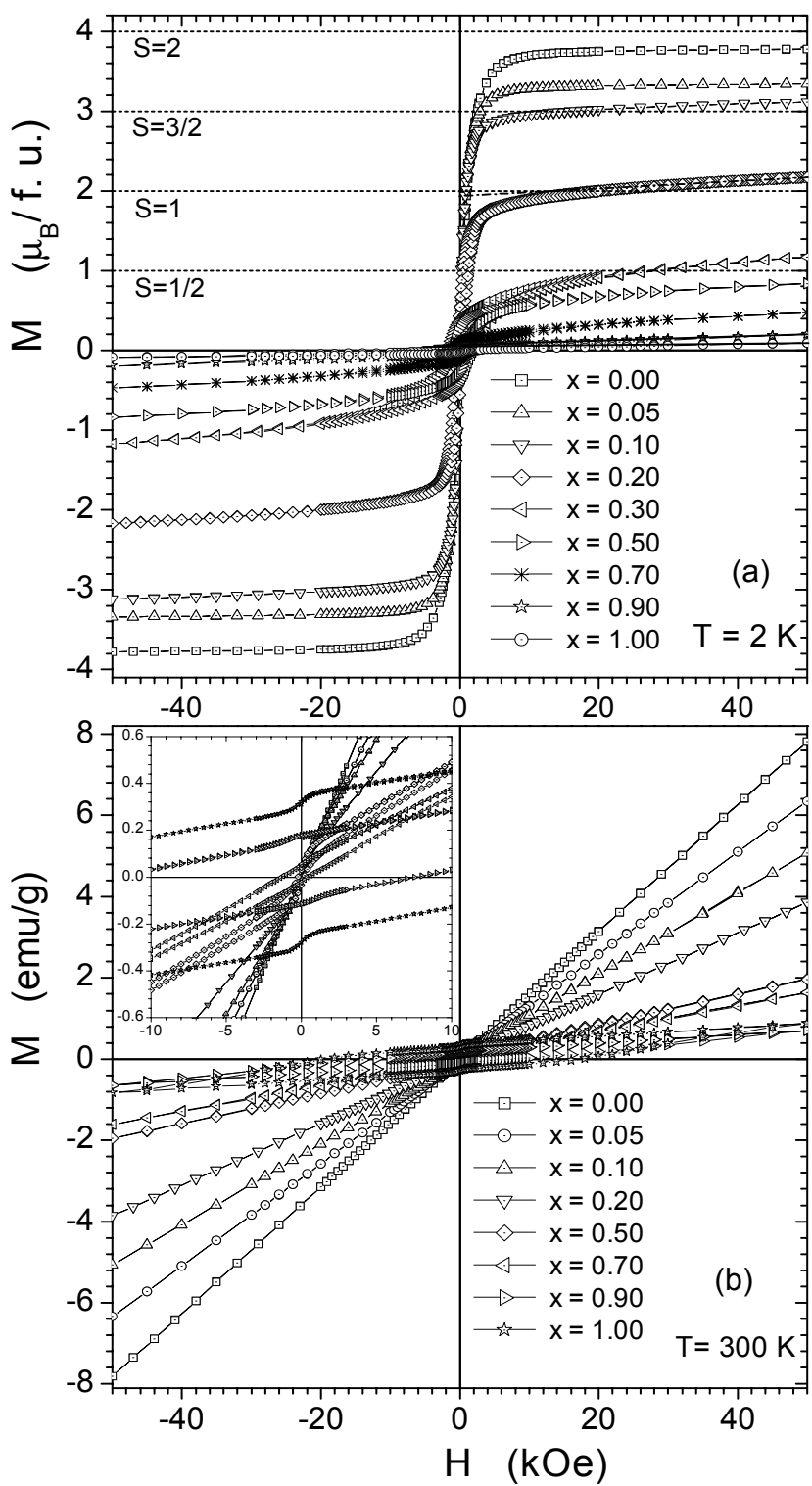

Fig.6 - O. F. de Lima et al.

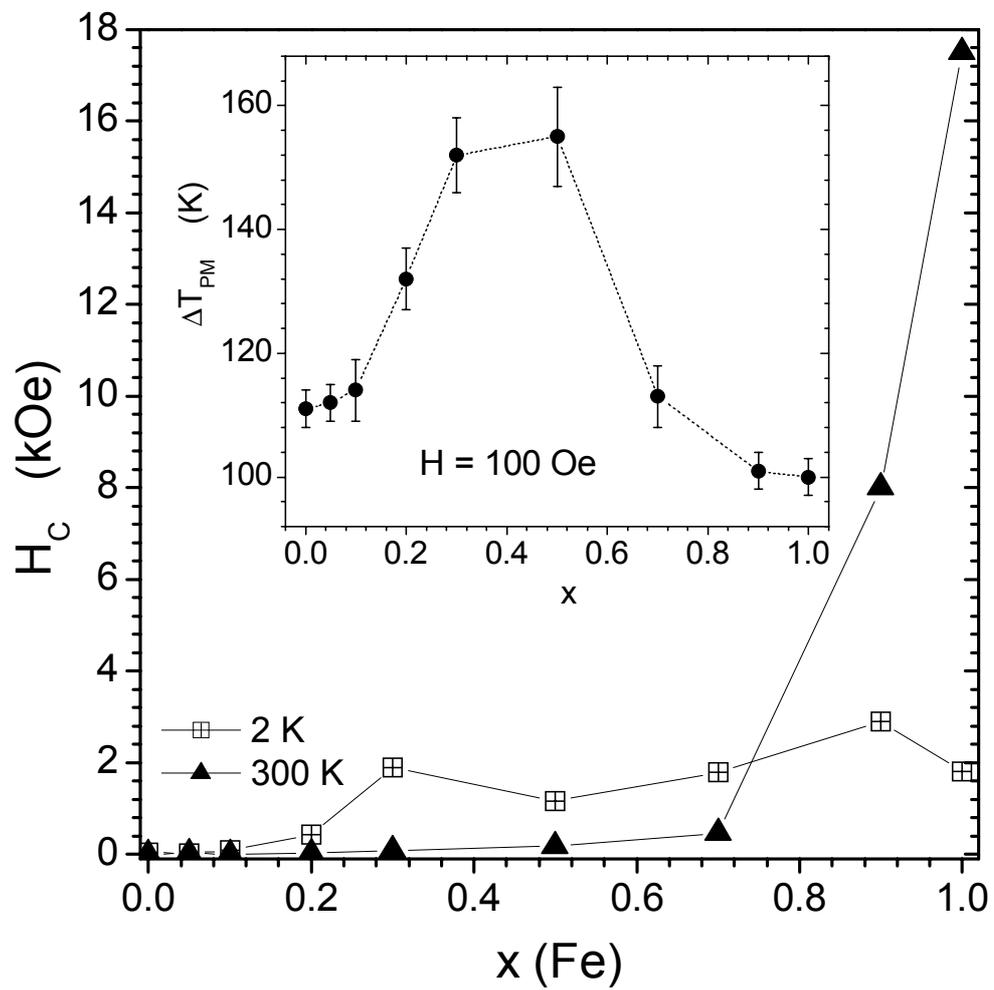

Fig.7 - O. F. de Lima et al.

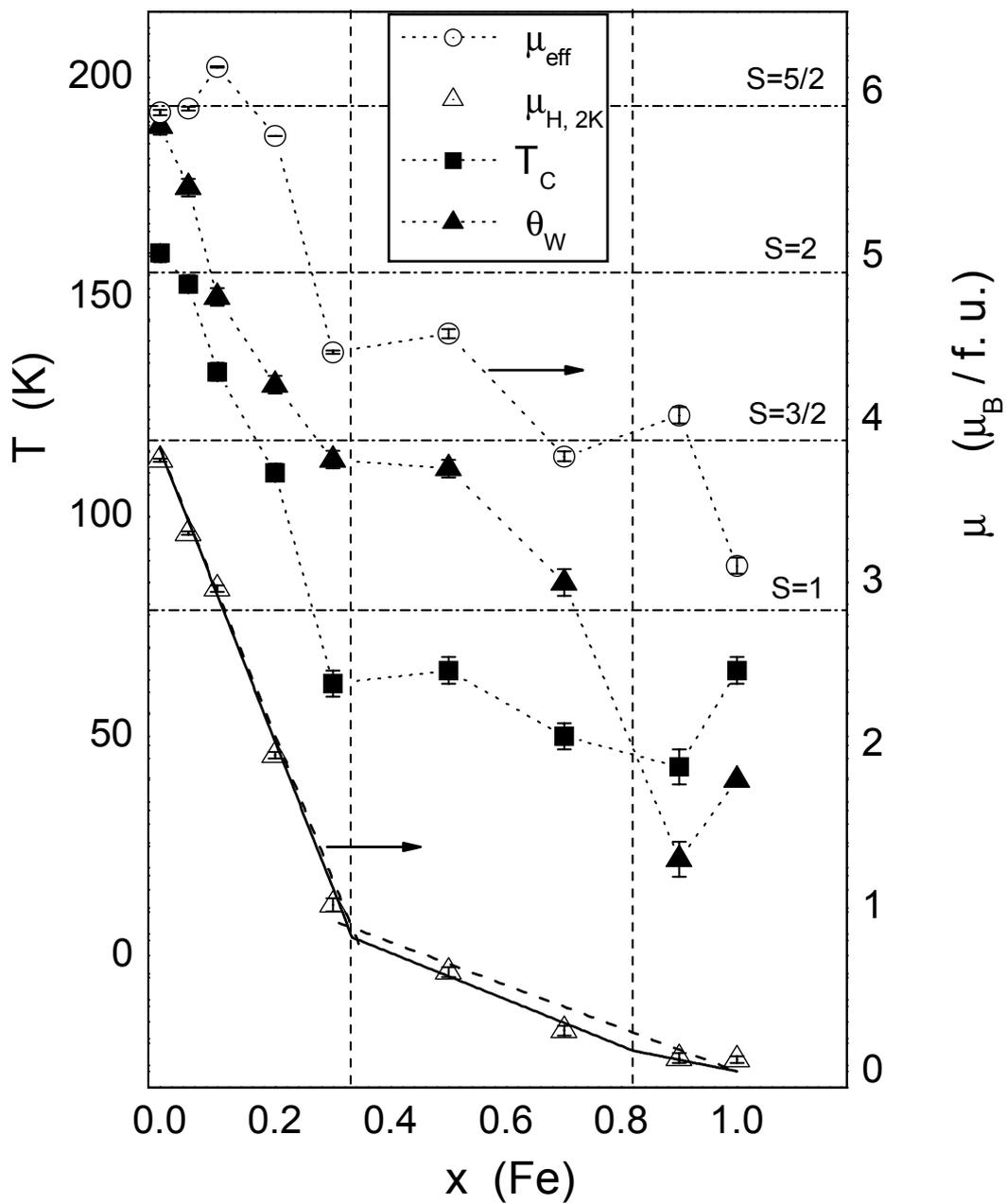

Fig.8 - O. F. de Lima et al.

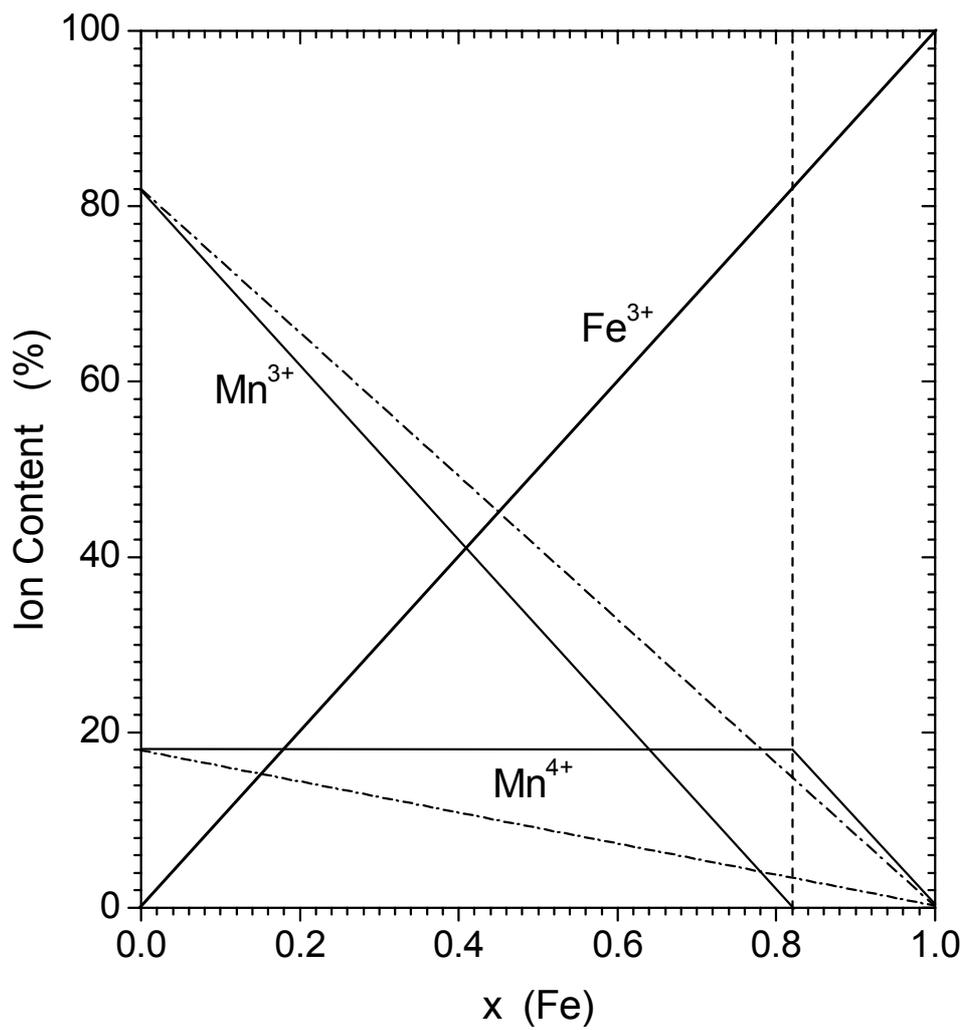

Fig.9 - O. F. de Lima et al.